\documentclass[12pt]{spieman}
\usepackage{amsmath,amsfonts,amssymb}
\usepackage{graphicx}
\usepackage{setspace}
\usepackage{caption}
\usepackage{subcaption}
\usepackage{hyperref}
\usepackage{hyperref}
\hypersetup{
    colorlinks=true,
    linkcolor=blue,
    filecolor=magenta,      
    urlcolor=blue,
    pdftitle={Overleaf Example},
    pdfpagemode=FullScreen,
    }
\usepackage[labelfont=bf]{caption}
\usepackage{inputenc}
\usepackage{parskip}
\usepackage{indentfirst} 
\usepackage{float} 

\DeclareCaptionFormat{myformat}{\fontsize{10pt}{10pt}\selectfont#1#2#3}
\usepackage{subcaption}
\usepackage{enumitem}
\usepackage{tikz}
\usepackage{float}
\usepackage{titlesec}
\usepackage{tikz,xcolor,hyperref}
\definecolor{lime}{HTML}{A6CE39}
\DeclareRobustCommand{\orcidicon}{
	\begin{tikzpicture}
	\draw[lime, fill=lime] (0,0) 
	circle [radius=0.16] 
	node[white] {{\fontfamily{qag}\selectfont \tiny ID}};
	\draw[white, fill=white] (-0.0625,0.095) 
	circle [radius=0.007];
	\end{tikzpicture}
	\hspace{-2mm}
}
\foreach \x in {A, ..., Z}{%
	\expandafter\xdef\csname orcid\x\endcsname{\noexpand\href{https://orcid.org/\csname orcidauthor\x\endcsname}{\noexpand\orcidicon}}
}

\usepackage{tocloft}

\title {}

\cftpagenumbersoff{figure}
\cftpagenumbersoff{table}

\begin{document} 
\begin{titlepage}

    \begin{center}
        \vspace*{3cm}
            
        \textbf{\fontsize{16pt}{19.2pt}\selectfont\bfseries
Identification of lung nodules CT scan using YOLOv5 based on convolution neural network}
            
        
        \normalsize


        \author[1,*]{Haytham Al Ewaidat\orcidA{}}
        \author[2]{Youness El Brag\orcidB{}}
        
        \affil[1]{Jordan University of Science and Technology, Faculty of Applied Medical Sciences, Department of Allied Medical Sciences-Radiologic Technology, Irbid, Jordan, 22110
        }
        \affil[2]{Abdelmalek Essaâdi University of Science and Technology, Faculty of  Multi-Disciplinary Larache, Department of Computer Sciences,  ksar el kebir , Morocco, 92150 } 
         

        \maketitle
            
       {\noindent \footnotesize\textbf{Correspondence author:} 
       \mdseries Dr Haytham Al Ewaidat, Department of Allied Medical Sciences-Radiologic Technology, Faculty of Applied Medical Sciences, Jordan University of Science and Technology. PO Box 3030, Irbid 22110, Jordan Tel: (+962)27201000-26939; Fax: (+962)27201087; E-mail: \linkable{haewaidat@just.edu.jo} }

    \end{center}
\end{titlepage}
\vspace*{5mm}

\begin{abstract}
\vspace{2mm}
\hfill \break 
\fontsize{10pt}{10pt} \textbf{ Purpose:} The lung nodules localization in CT scan images is the most difficult task due to the complexity of the arbitrariness of shape, size, and texture of lung nodules. This is a challenge to be faced when coming to developing different solutions to improve detection systems. the deep learning approach showed promising results by using convolutional neural network (CNN), especially for image recognition and it's one of the most used algorithm in computer vision.
\newline
\fontsize{10pt}{10pt} \textbf{Approach:}  we use (CNN) building blocks based on YOLOv5 (you only look once) to learn the features representations for nodule detection labels, in this paper, we introduce a method for detecting lung cancer localization. Chest X-rays and low-dose computed tomography are also possible screening methods, When it comes to recognizing nodules in radiography, computer-aided diagnostic (CAD) system based on (CNN) have demonstrated their worth. One-stage detector YOLOv5 trained on 280 annotated CT SCAN from a public dataset LIDC-IDRI based on segmented pulmonary nodules. 
\newline
\fontsize{10pt}{10pt} \textbf{Results}: we analyze the predictions performance of the lung nodule locations, and demarcates the relevant CT scan regions. In lung nodule localization the accuracy is measured as mean average precision (mAP). the mAP takes into account how well the bounding boxes are fitting the labels as well as how accurate the predicted classes for those bounding boxes, the accuracy we got 92.27\% .
\newline
\fontsize{10pt}{10pt} \textbf{Conclusion:} this study was to identify the nodule that were developing in the lungs of the participants. It was difficult to find information on lung nodules in medical literature, 
\end{abstract}
\keywords{computer-aided diagnostic, deep learning,  Convolutional Neural Networks ,Lung Nodule}
\newline
{\noindent \footnotesize\textbf{*}Address all correspondence to Haytham Al Ewaidat ,  \linkable{haewaidat@just.edu.jo} }

\section{introduction}
As far as noninvasive therapy and clinical assessment are concerned, medical image analysis offers a tremendous advantage. X-rays, CTs, MRIs, and ultrasounds are utilized to make precise diagnoses based on the obtained restorative images. By using attractive fields, CT can capture pictures on film in medical imaging. One-of-a-kind lung cancer is responsible for 1.61 million fatalities per year. Most of the cases of lung cancer in Indonesia are observed in the MIoT centers. If the tumor is identified early, the survival percentage is better then. It's not an easy task to find lung cancer in its early stages.
Approximately 80\% of cancer patients are diagnosed at the core or accelerated phase of the disease. Lung cancer is the second most common cancer among men and the tenth most common among women worldwide. After breast and colorectal cancer, lung cancer is the thirdly most common cancer among women. Features extraction in image processing is one of the simplest and most efficient dimensionality reduction approaches. The non-invasive nature of CT imaging is one of its most notable characteristics. It's surprising to see angles increasing when compared to other imaging modalities.

Computed tomography imaging is the best technique for examining lung disorders. CT scans, on the other hand, have a high probability of false-positive results and are associated with cancer-causing radiation exposure. When compared to standard-dose CT, low-dose CT utilizes a lot less radiation contact power. The findings reveal that the detection sensitivity of low-dose and standard-dose CT images is not significantly different. A well know database the National Lung Screening Trial database shows that cancer-related fatalities were considerably decreased in the group that was subjected to low-dose CT scans rather than chest radiography. The sensitivity of lung nodule identification may be improved by the use of more detailed anatomical information, and better image registration methods. As a result, the datasets have grown enormously. Up to 500 segments/slice may be generated from a single scan, depending on how thick the slice is. A single slice is examined by a competent radiologist in 2–3.5 minutes. A radiologist's workload keeps rising while screening a Ct for the presence of a suspicious nodule. The detection sensitivity of nodules is influenced by a variety of factors, including the size, location, form, nearby structures, edges, and density, in addition to the CT slice section thickness.

Only 68 percent of lung cancer nodules are properly identified when only one radiologist doctor views the scan, and up to 82\% of the time when two radiologists check the scan, according to the study results. Early diagnosis of malignant lung nodules by radiologists is a tough, time-consuming, and laborious process in and of itself. The radiologist needs a lot of time to carefully screen a large number of images, but this method is prone to mistake when looking for microscopic nodules.

An aid for radiologists is required in this case to speed up readings, catch any missing nodules, and enable improved localization. A primary goal of computer-aided detection systems was to minimize radiologists' labor and boost the detection rate of nodules. Newer CAD systems, on the other hand, can distinguish between benign, and malignant nodules, which is helpful in the screening process. CAD systems frequently beat professional radiologists in nodule identification and localization tasks because of recent breakthroughs in deep learning models, particularly in image processing. CAD systems, on the other hand, have an FP rate of 1–8.2 per scan and a detection range of 38–100\%, according to different studies. As a result of their likeness to one other, benign, and malignant nodule remain a difficult challenge to solve.

During the screening process, a variety of mistakes might occur. For example, if a scan fails to capture or recognize a specific region of the lesion or fails to distinguish between benign, and malignant lesions in a patient's body, the patient may be at risk of misdiagnosis. Most people die as a result of misdiagnoses and delays in treatment because of these mistakes. In radiology, over 4\% of reports include diagnostic mistakes on a daily, and about 30\% of aberrant radiological diagnoses are ignored. Early-stage lung nodules may be detected and classified more accurately using different methodologies such as deep learning.

Lung nodule identification using deep learning with a specific methodology is presented in this research. lung CT images, physiological symptoms, and clinical indicators, the suggested approach reduces false-positive findings and eventually prevents invasive procedures. YOLOv5 is used which has convolutional networks were built to identify and classify nodules. For nodule identification. Nodule identification and classification using the publicly accessible data set LIDC-IDRI surpasses state-of-the-art deep learning techniques. Using a variety of techniques, we were able to reduce the number of false positives in the learning algorithm.

Lung nodule computer-aided detection (CAD) systems were originally developed in the late 1980s, but the processing resources required for sophisticated image analysis methods at the time made these efforts unattractive. For image analysis, and decision support systems based on computers, the graphics processing unit and convolutional neural networks revolutionized their performance. Some of the most important lung nodule identification and classification approaches have been suggested by researchers in deep learning based medical images analysis models. For lung nodule classification, Yutong Xie et al \cite{r1} . proposed a method that utilizes Texture, Shape, and Deep Model learned Data at the choice level.

Nodule heterogeneity may be shown with the use of this algorithm's GLCM-based surface descriptor, Fourier-shape descriptor, and a DCNN. Based on CNNs Chougrad et al \cite{r2}. studied the classification of breast cancer using a CAD framework. Transfer learning, on the other hand, takes just a small number of medical images to train a system. With the use of the transfer learning approach, the CNNs were taught to their fullest potential. In terms of accuracy, CNN came out on top with a score of 98.94 percent. Using the wavelet transform and principal component analysis, Heba Mohsen et al\cite{r3} developed a DNN classifier for brain tumor classification. A technique of regularized linear discriminant analysis was developed in 2015 by Sharma et al \cite{r4}, and it used a regularization parameter to perform a standard cross-validation methodology. An appropriate collection of characteristics is needed to evaluate medical data for illness prediction. Several evolutionary algorithms have been used to find the best possible traits. Gravitational search and Elephant Herd optimizations have recently been used to choose the best features \cite{r5}. Another deep learning-based model created by Kuruvilla, and Gunavathi, K. in 2014, an ANN-based cancer classification for CT scans. Development of the statistical model used to classify the data was completed. Compared to feed-forward networks feed-forward backpropagation networks are more accurate, according to research. Classifier accuracy may be improved even more by using the skewness feature \cite{r6}.

Lung cancer detection categorization is becoming more and more popular due to the rapid advancement of pattern recognition and image processing methods. Textural evaluation of thin-section CT images has been used in the literature to help distinguish various obstructive lung disorders. Attenuation distribution statistics, acquisition-length parameter, and co-occurrence descriptor are all included in 13-dimensional vectors of local textures information developed by Chabat et al\cite{r7}. A Bayesian classifier is used for feature segmentation. These five scalar metrics, max, entropy, energy, contrast, and homogeneity were recovered per each co-occurrence matrix to minimize the feature vector's dimensionality. The textural characteristics of Solitary Pulmonary Nodules discovered by CT have been described and assessed by Yanjie Zhu et al \cite{r8}. It took 300 generations for 67 characteristics to be retrieved, however, only 25 features were picked. SVM-based classifiers are used for classification. For Interstitial Lung Disease, Sang Cheol Park and colleagues \cite{r9} used a genetic algorithm to identify the best picture attributes (ILD). Hiram et al \cite{r10} used the frequency domain, and SVM with RBF to classify lung nodule classifications. Solitary pulmonary nodules may be automatically detected using an algorithm provided by Hong et al\cite{r11}. True nodules are identified and labeled on original images using an SVM classifier. The LIDC-IDRI images database was used by Antonio et al\cite{r12}, to classify lung nodules.  Ecological taxonomic diversity and taxonomic distinctness indexes are used for classification using SVM \cite{r13}. Results show a 98.11\% accuracy rate. 

The mesh grid region growth approach was used in CT to select and analyze just the pixels that were most likely to be relevant to the diagnosis. The ILD status of all unselected pixels was determined to be negative. To recognize lung cancer cells, Zhi-Hua et al\cite{r14} presented Neural Ensemble-based Detection (NED), which makes use of an artificial neural network ensemble. Using this technology, it is possible to accurately identify cancer cells. An algorithm developed by Hui Chen et al\cite{r15}, uses a Neural Networks Ensemble to construct the categorization of a lung nodule on a thin section CT image (NNE). A model suggested by Aggarwal, Furquan, and Kalra \cite{r16} is characterized by normal lung architecture by which segmentation is done using the best possible thresholds. Geometric, statistical, and grey level properties are used to extract features. Classification is done using LDA. The accuracy is 84\%, the sensitivity is 97\%, and the specificity is 53\%. An inference-based approach to identify lung cancer nodules has been developed by Roy, Sirohi, and Patle \cite{r17}. To improve contrast, this technique employs grey transformations using an active contour model, the image is segmented. Training the classifier is done by extracting features such as area, mean, major axis, and minor axis length. Overall, the system's accuracy rate is 94.12\%. This approach has a disadvantage in that it does not distinguish between benign cancers and those that are malignant. Authors have used wavelet feature descriptors to classify lung nodules \cite{r18}. One and two-level decompositions of wavelet transformations are used in this example. A total of 19 characteristics are derived from each wavelet sub-band. SVM is used to distinguish between CT images that include malignant nodules and those that do not.
\section{\fontsize{12pt}{12pt}Material and Methods} 
In this section, we introduce our methods for Lung Nodules localization  We use a one-stage-method based on YOLOv5 detection , the methodology has been split into the following Subsections to explain the whole process of our method .
\hfill
\subsection{ Dataset}
For this research, the dataset has been collected from LIDC-IDRI. In LIDC-IDRI image collection, thoracic CT scans with marked-up annotated lesions are included. For the development, training, and assessment of computer-assisted diagnostic (CAD) approaches for the detection and diagnosis of lung cancer is a worldwide web-accessible resource One example of a public-private partnership founded on consensus-based decision-making is this collaboration between the National Cancer Institute, the Foundation for the National Institute of Health, and Food and Drug Administration (FDA), which was spearheaded by NCI and supported by the FDA. This data collection, which includes 239 Ct images for training and 41 images for validation. is a subset of the original dataset. Some of the samples are given below in the following  Fig.\ref{fig:samples}.
\begin{figure}[H]
    \centering
    \includegraphics[width=0.3\textwidth] {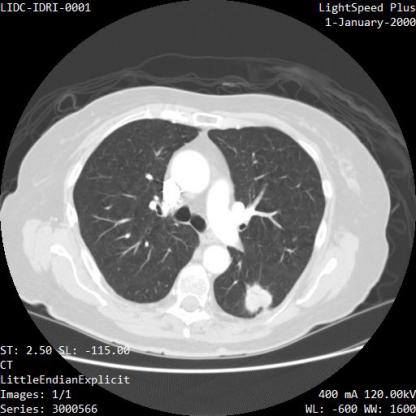}
    \includegraphics[width=0.3\textwidth]{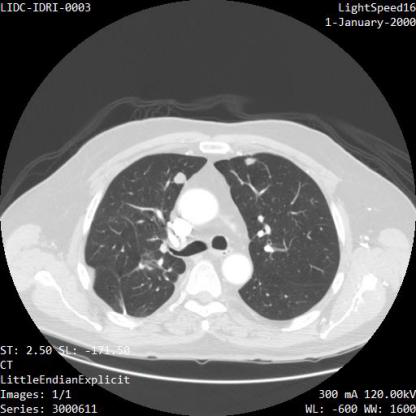}
    \vspace{1.5ex}
    \caption{Samples from dataset LIDC-IDRI Lung Cancer }
    \label{fig:samples}
\end{figure}
 
\subsection{\fontsize{11pt.0}{12pt}  Pre-Processing Data }
Real-world data tends to be fragmentary, noisy, and inconclusive. This may lead to low-quality data collection, which in turn can lead to low-quality models. Data Preprocessing offers procedures that may properly organize the data for better comprehension in the deep learning process to solve these challenges. Data Preprocessing steps that have been used in this research study are given in the following Fig.\ref{fig:process}.\hspace*{\fill}
\begin{figure}[H]
    \centering
    \includegraphics[width=0.6\linewidth]{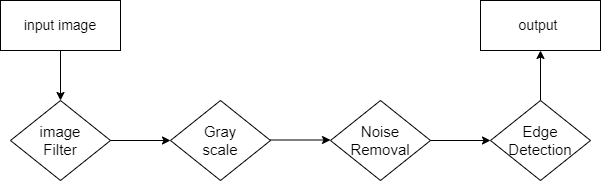}
    \vspace{1.5ex}
    \caption{Preprocessing steps for images}
    \label{fig:process}
\end{figure}    

\subsection{\fontsize{12.0pt}{12pt}\selectfont Model architecture }
As discussed in the introduction in this research YOLOv5 model is used for feature extraction and detection of lung nodules in CT scans. Let have a brief discussion about Yolov5 and its architecture. 
\subsubsection{\fontsize{11.0}{12pt}\selectfont YOLOv5 for lung nodules localization }
\begin{figure}[H]
    \centering
    \includegraphics[width=\linewidth]{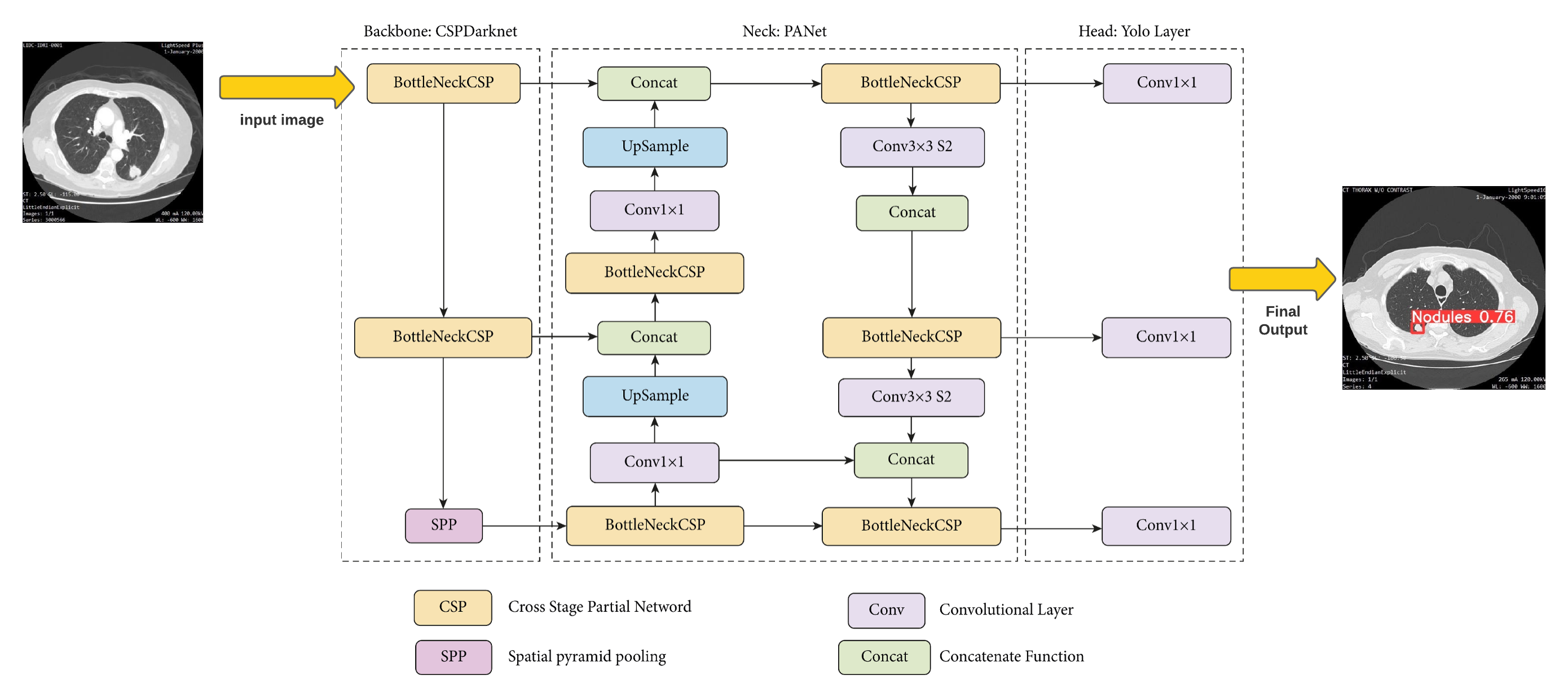}
     \vspace{1.5ex}
    \caption{Overview of YOLOv5 building blocks model architecture
     \label{marker}}
    \label{fig:yolo}
\end{figure}

\begin{figure}[H]
    \centering
    \includegraphics[width=\linewidth]{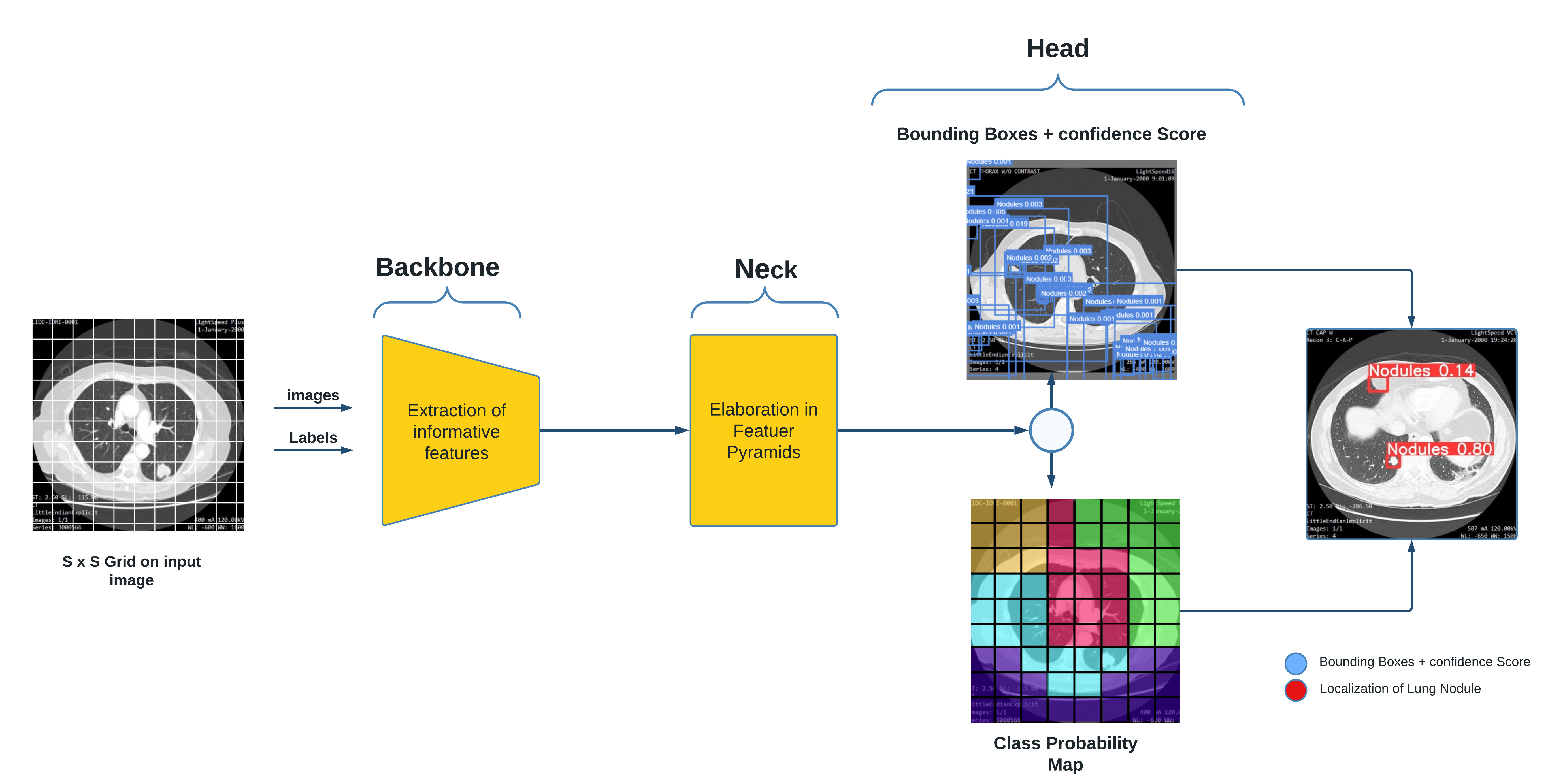}
     \vspace{1.5ex}
    \caption{ Model detection can be considered a regression problem. The image is divided into S * S grids in which bounding boxes are predicted for each grid cell, along with their confidence value }
     \label{fig:yolo_v5}
\end{figure}

the whole structure of Yolov4 Optimal speed and accuracy of object detection \cite{r19} is shown in Fig.\ref{fig:yolo} and YOLOv5 illustration representation shown in Fig.\ref{fig:yolo_v5}.
the YOLO family of models consists of three main components to every single-stage object detector, and YOLOv5 has its own three main modules\vspace{1.5ex}
\begin{enumerate}[label=(\arabic*)]
    \item Backbone \autoref{marker}:it's mostly used to extract the elements of the most significant feature from the images that have been provided. Cross Stage Partial Networks(CSP) is the backbone of YOLOv5's feature extraction, which uses them to extract an image's most informative details
    \item Neck \autoref{marker}: it used to create feature pyramids. Feature pyramids aid models in generalizing successfully when it comes to object scaling. It aids in the identification of the same item at various scales and dimensions.
    Feature pyramids are quite valuable and can help models perform effectively on data that has never been examined. It's not only FPN, BiFPN, and PANet that are used in feature pyramid models
    \item Head \autoref{marker}: it has layers that generate predictions from anchor boxes on features and generated final output vectors with probabilities, object classes scores, and bounding boxes., YOLOv5 uses the following choices for training \cite{r20} 
\end{enumerate}
\vspace{0.3cm}
\subsubsection{\fontsize{11.0}{12pt}\selectfont  Training Model }
\begin{figure}[H]
    \centering
    \includegraphics[width=0.6\linewidth]{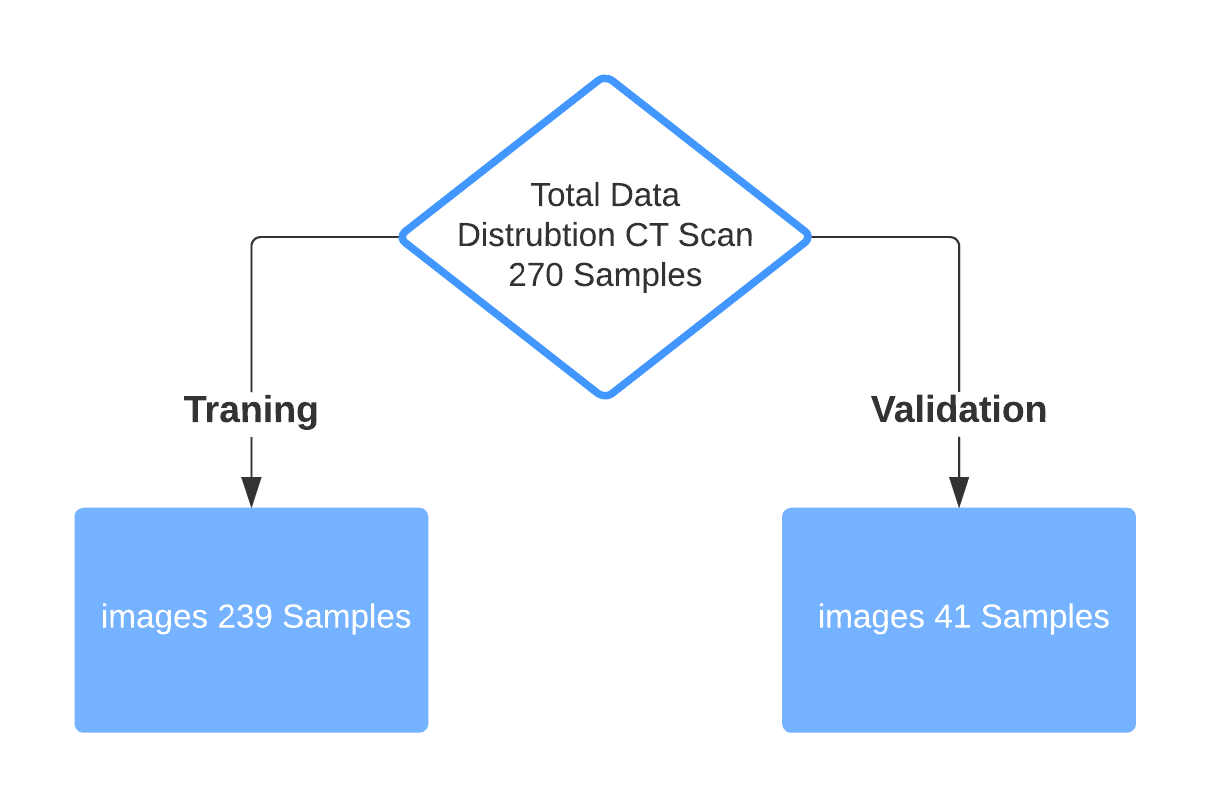}
    \vspace{1.5ex}
    \caption{ Dataset Splitting Diagram CT Scan images}
    \label{fig:Diagram}
\end{figure}

During the training and validation process, a total of 270 CT Scan images are used of which 239 CT Scans are used for training and 41 are used for validations. For training, the Google Colab is used which is an online platform for training models. Which provides 16GB GPU free for training. The batch size was kept to 16 and the number of epochs was kept to 100. Splitting of data can be seen in  Fig.\ref{fig:Diagram} .
\nopagebreak
\section{Results} 
the model had initial leverage to train faster and predict the location of lung nodules and demarcates the relevant CT scan regions. before diving into the analysis of the results is necessary to explain the statistical machine learning  knowledge behind those results, the explanations have been split into the following Subsections to explain the whole analysis of the method we use. 
\subsection{\fontsize{11.0}{12pt}\selectfont Evaluation Metrics }
In this section, we describe Charts of evaluation metrics that got from our experiment. It is known to us that, in the computer Aide system, the main part is detecting the object inside the image. Common metrics for measuring the performance of classification algorithms such as YOLOv5 that are based on CNN include, Recall, precision, F-score, mAP, PR\_curve, F1\_curve
, IOU \cite{r22}, overlapping error, and boundary-based evaluation, the evaluation metrics we used is the mean Average Precision (mAP) \cite{r23}, 
the precision, and F1-Curve. We will briefly explain them in the following part. According to the theory of the statistical machine learning , precision is a two-category statistical indicator whose formula is .\vspace{1.5ex}

Precision : measures how accurate is our predictions was. the percentage of our predictions are correct as shown in Fig.\ref{fig:Evaluation},and following equation\ref{eq:1}.
\begin{center}
\begin{equation} \label{eq:1}
 Precision = \frac{TP}{TP+FP}
\end{equation}
\end{center}
\begin{center}
\end{center}
 Recall: measures how much of the true bbox were correctly predicted as shown in the following equation.\ref{eq:2}.
\begin{center}
\begin{equation}\label{eq:2}
     Recall = \frac{TP}{TP+FN}
\end{equation}
\end{center}
moreover, it is necessary to know TP, FP, and FN in the localization Nodules task.\hspace*{\fill}
\begin{enumerate}[label=(\arabic*)]
    \item True positive (TP): IoU$>$[formula] (in this work, [formula] takes 0.2) the number of Localization frames (the same Ground Truth is only calculated once\cite{r21}) 
    \item
    False positive (FP): the number of check boxes for IoU$<=$[formula] or the number of redundant check boxes that detect the same Ground Truth
    \item  False negative (FN): the number of Ground Truths not detected 
\end{enumerate} 
the IoU is a  measures of the degree of overlap between two  boundaries. We use that to measure how much our predicted frame  overlaps with the ground truth (the actual ground frame) ,the IOU is shown with Fig.\ref{fig:Predicted} as follows, and the formula is as following Fig.\ref{fig:formula}:
\begin{center}
\begin{figure}[H]
    \centering
    \includegraphics[width=0.9\linewidth]{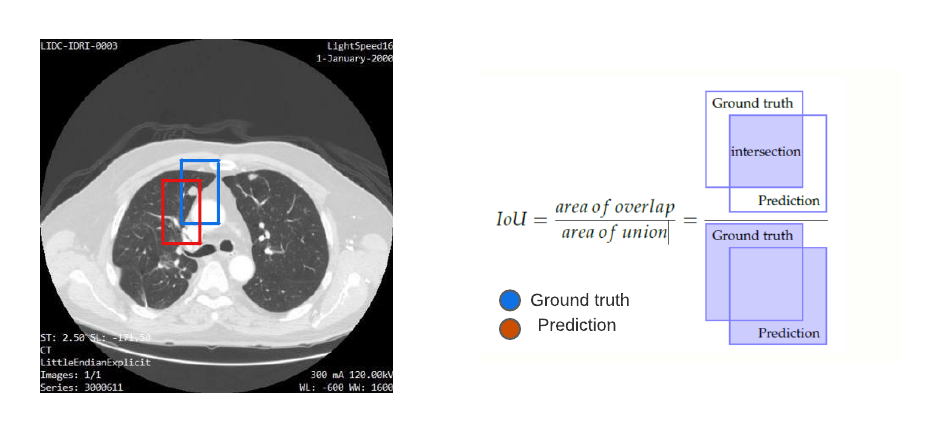}
    \caption{Graphical representation of the Intersection over Union (IoU=0.2) calculation on a narrow-band imaging. The light blue rectangle
    represents the ground truth bounding box, while the red rectangle represents the model prediction. The IoU is calculated by dividing the overlap area by the total area of union}
    \label{fig:formula}
\end{figure} 
\end{center}
after getting familiar with these definitions of statistical learning formulas, we introduce the mAP (mean Average Precision). The mAP compares the ground-truth bounding box to the detected box and returns a score. The higher the score, the more accurate the model is in its detections.\\F1-score is defined as the harmonic average of precision and recall as shown in figure \ref{fig:F_score}:
\begin{center}
\begin{equation}
    F1\_Score = \frac{2*Precision*Recall}{Precision+Recall}
\end{equation}
\end{center}
\begin{figure}[H]
     \centering
\begin{minipage}[b]{0.58\textwidth} 
\begin{subfigure}{0.49\textwidth}
         \centering
         \includegraphics[width=1\linewidth]{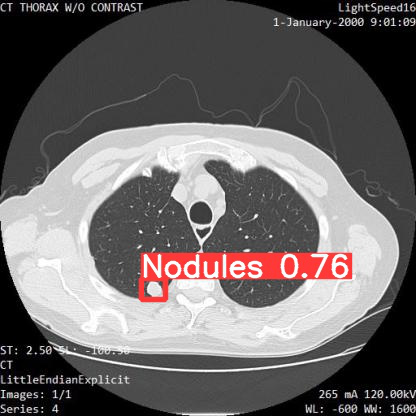}
         \caption{the Predicted location of a Single Nodule}
\end{subfigure}\hfill%
\begin{subfigure}{0.49\textwidth}
 \centering
         \includegraphics[width=\textwidth]{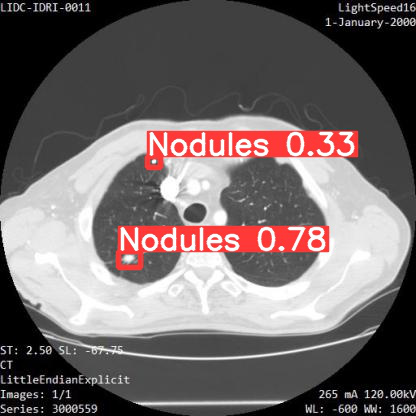}
         \caption{the Predicted location of tow Nodules in region }
\end{subfigure}
\end{minipage}
   \vspace{1.5ex}
   \caption{Example of output Results}
   \label{fig:Predicted}
\end{figure}
\subsection{\fontsize{11.0}{12pt}\selectfont  Experiment’s Setting }
the set Hyper-parameters of our fine tuning model are shown in Table 1, Our experiment uses Pytorch framework deep learning on GPU Tesla K80 by Google open Platform Colab-research .
\begin{center}
\begin{table}[htp]
\centering 
\setlength{\tabcolsep}{0.67cm} 
\renewcommand{\arraystretch}{1.2}
\begin{tabular}{c c c c} 
\hline 
Parameters &  &  & Value \\ [0.5ex] 
\hline 
Batch size &  &  & 16 \\ 
Image size &  &  & 416 \\
Epoch &  &  & 145 \\
Learning rate &  &  & 0.01 \\
Optimizer  &  &  & SGD \\ [1ex] 
\hline 
\end{tabular}
\vspace{1.5ex}
\caption{Parameters and their value.\\}
\end{table}
\end{center}
\subsection{\fontsize{11.0}{12pt}\selectfont Experiment’s Result and Analysis }
\begin{figure}[!ht]
    \centering
    \begin{subfigure}{.4\linewidth}      
        \centering
        \includegraphics[width = \linewidth]{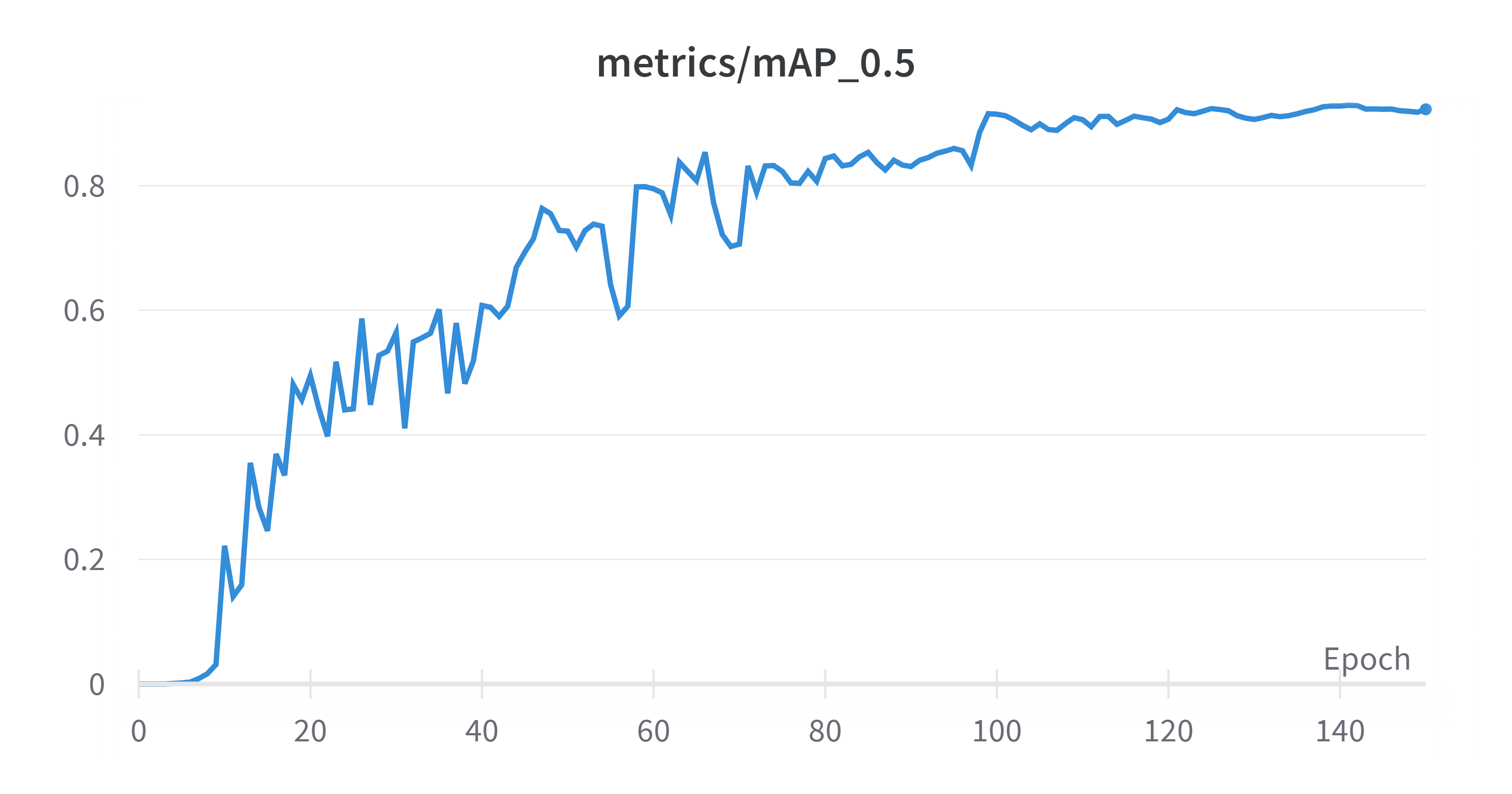}
         \caption{mean Average Precision Evaluation}
         \label{fig:A}
    \end{subfigure}
    \begin{subfigure}{.4\linewidth}   
        \centering
        \includegraphics[width = \linewidth]{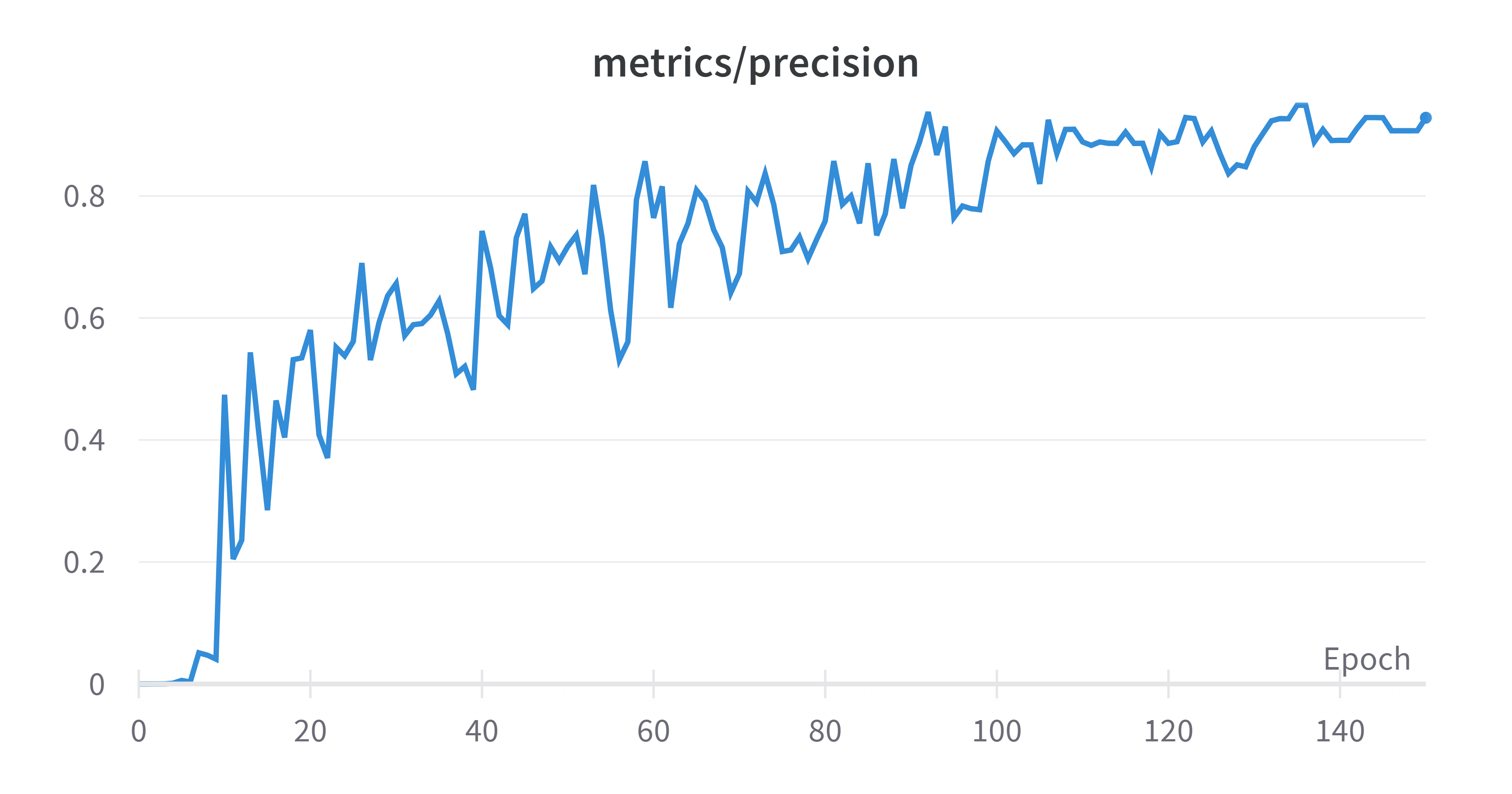}
         \caption{Precision Evaluation }
         \label{fig:B}
    \end{subfigure}
    \vspace{1.52ex}
   \caption{the important Evaluation Metrics}
  \label{fig:Evaluation}
\end{figure}

\begin{figure}[!ht]
  \centering
  \begin{subfigure}{.4\linewidth}
    \centering
    \includegraphics[width = \linewidth]{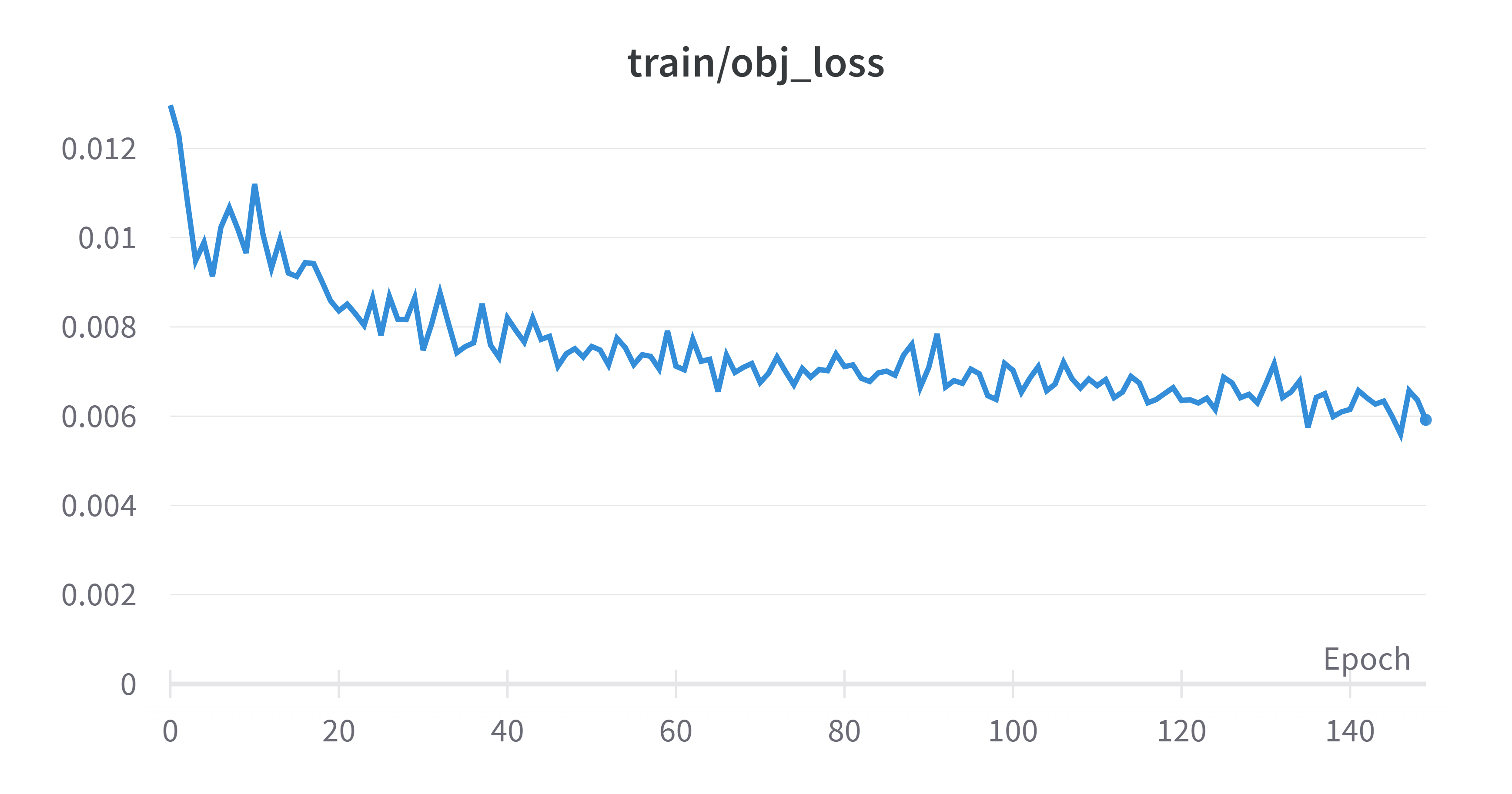}
    \caption{ the training confidence of object presence loss}
  \end{subfigure}%
  \hspace{1em}
  \begin{subfigure}{.4\linewidth}
    \centering
    \includegraphics[width = \linewidth]{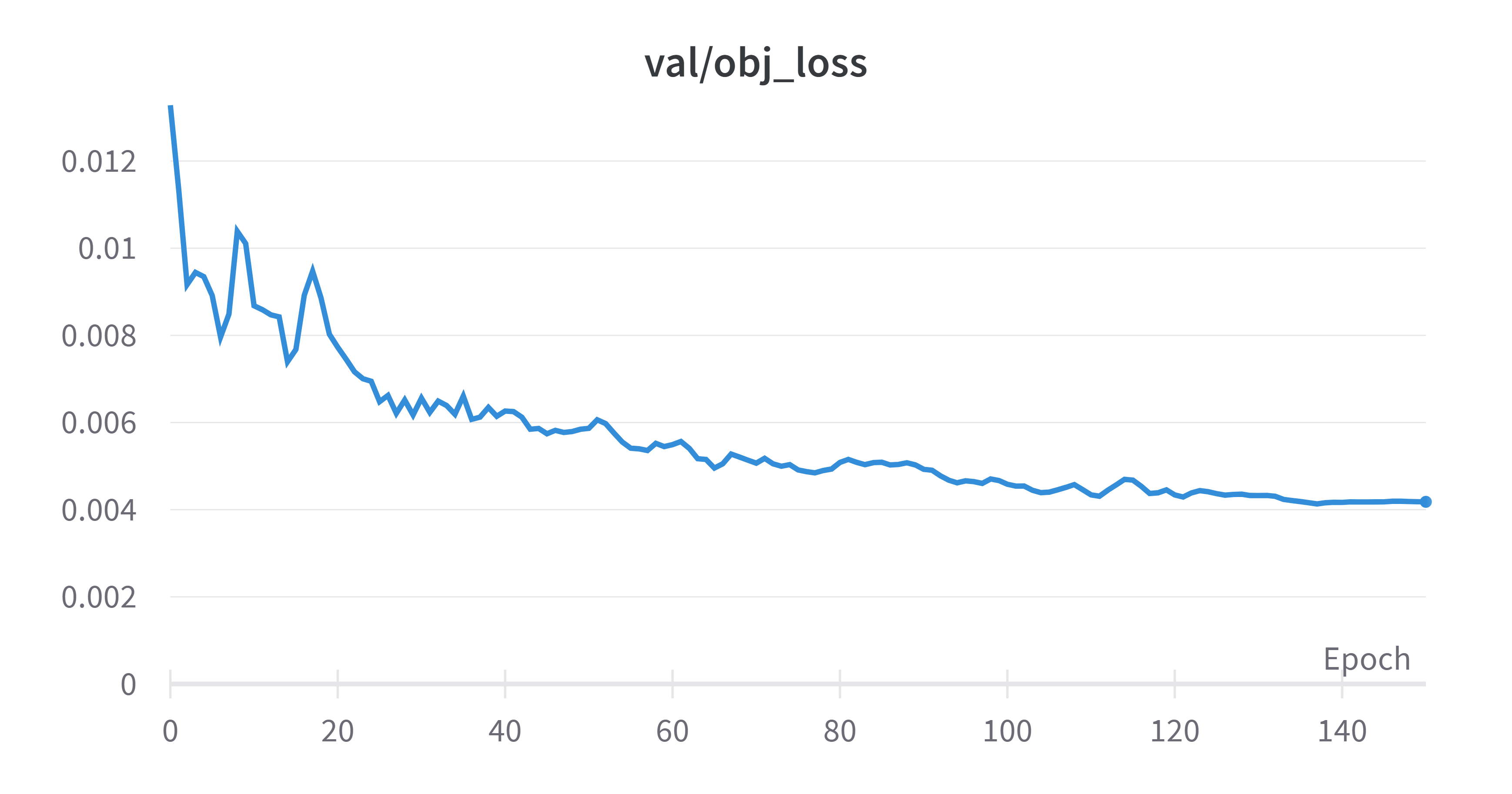}
    \caption{the validation confidence of object presence loss }
  \end{subfigure}%
  \hspace{2em}
  \begin{subfigure}{.4\linewidth}
    \centering
    \includegraphics[width = \linewidth]{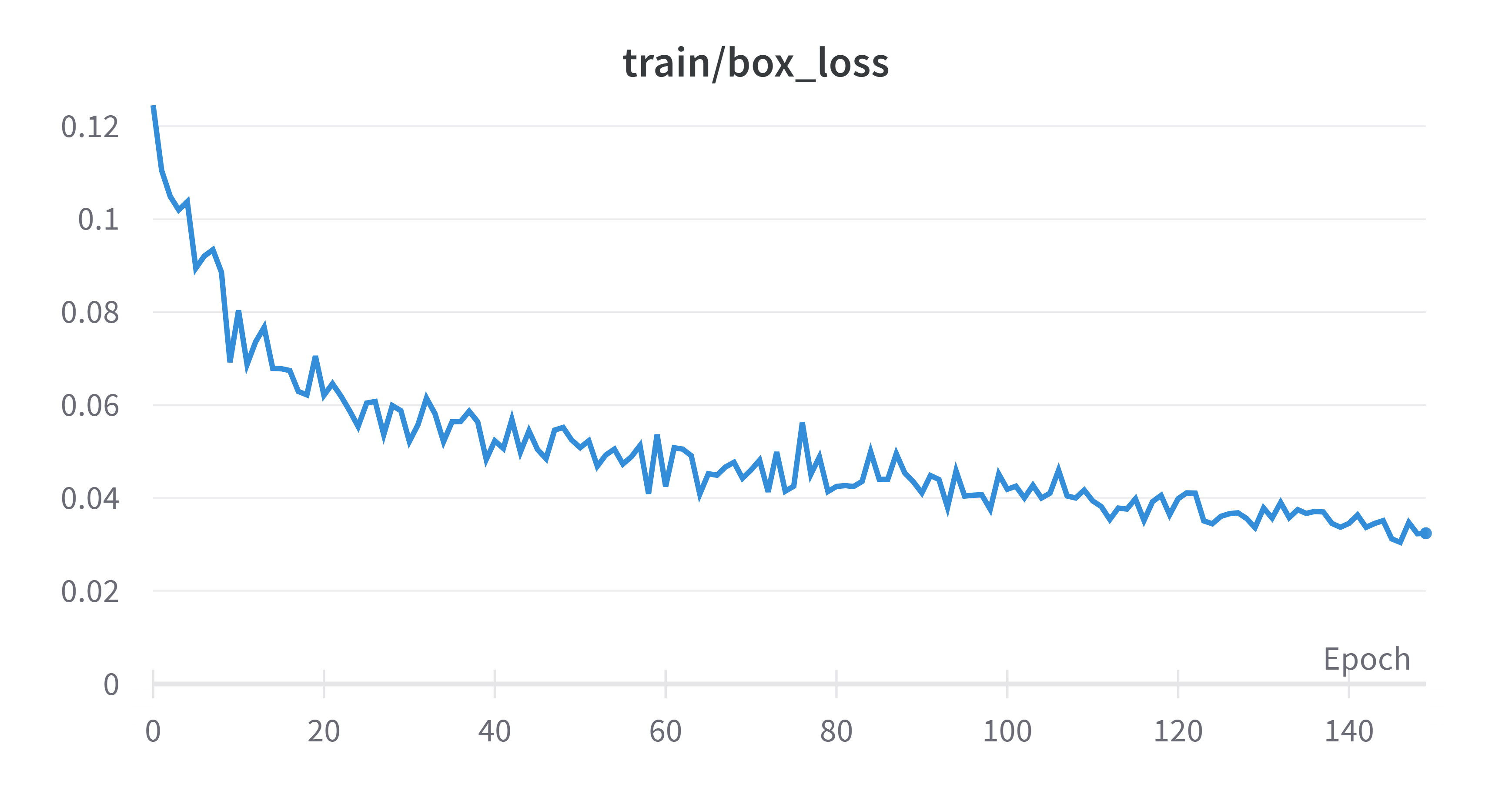}
    \caption{the training bounding box regression loss }
  \end{subfigure}
  \begin{subfigure}{.4\linewidth}
    \centering
    \includegraphics[width = \linewidth]{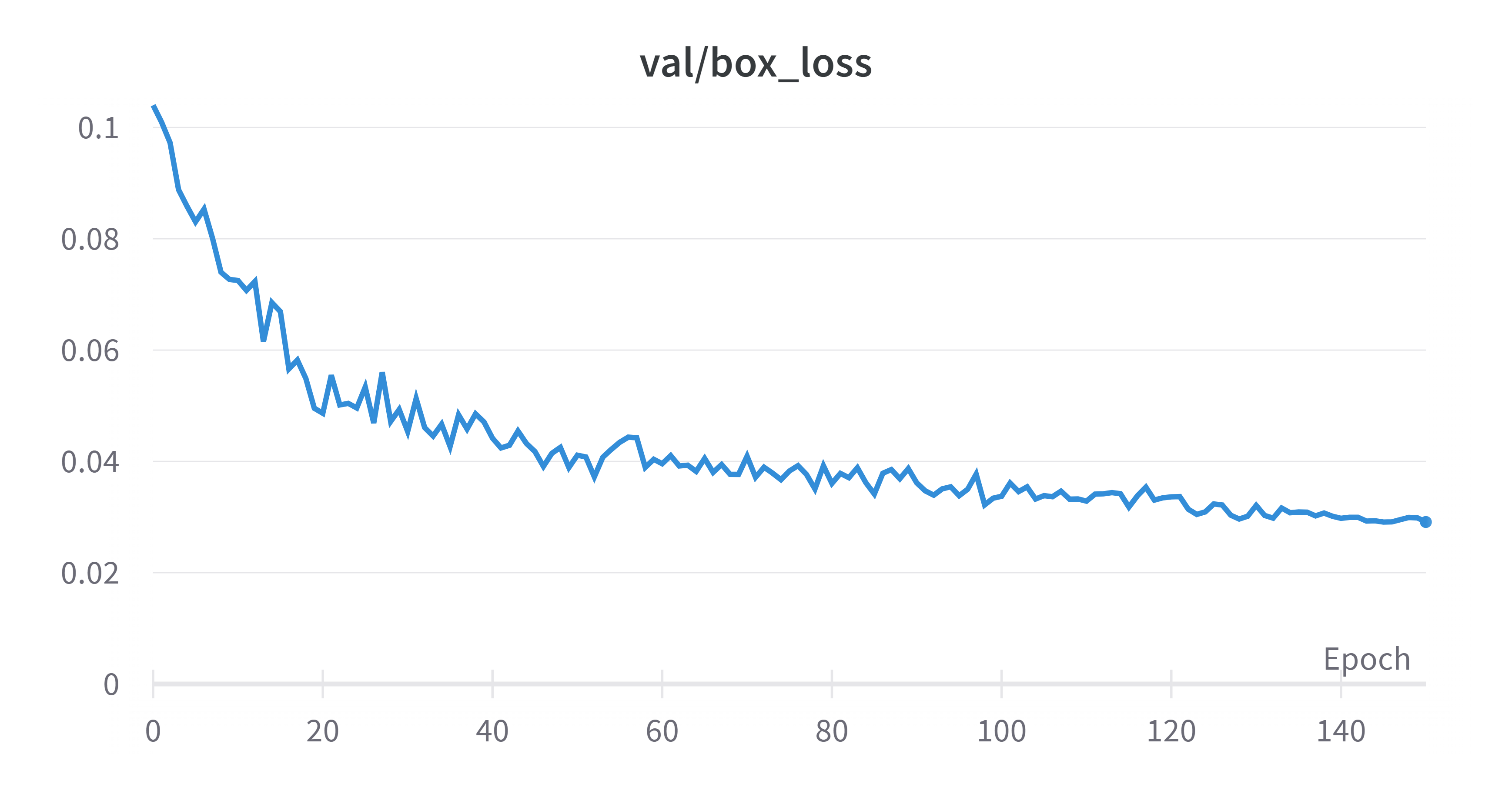}
    \caption{the validation bounding box regression loss}
  \end{subfigure}
  \vspace{1.52ex}
  \caption{Results of feature extraction training and validation }
  \label{fig:plot_tranign}
\end{figure}

\begin{figure}[!ht]
    \centering
    \begin{subfigure}{.4\linewidth}      
        \centering
        \includegraphics[width=\linewidth]{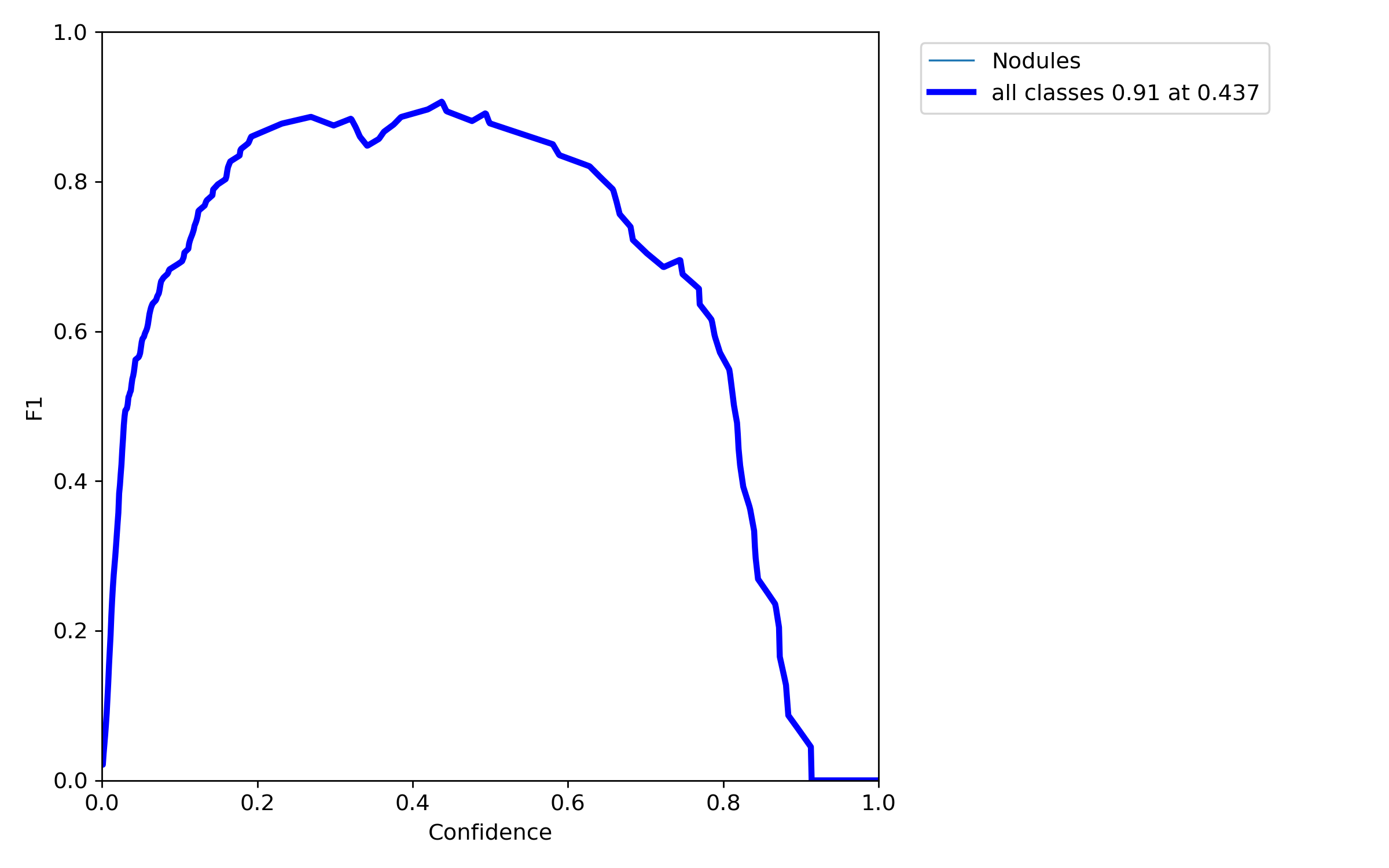}
         \caption{F1 indicator training process for single category}
         \label{fig:F_score}
    \end{subfigure}\hfill%
    \begin{subfigure}{.4\linewidth}   
        \centering
        \includegraphics[width=\linewidth]{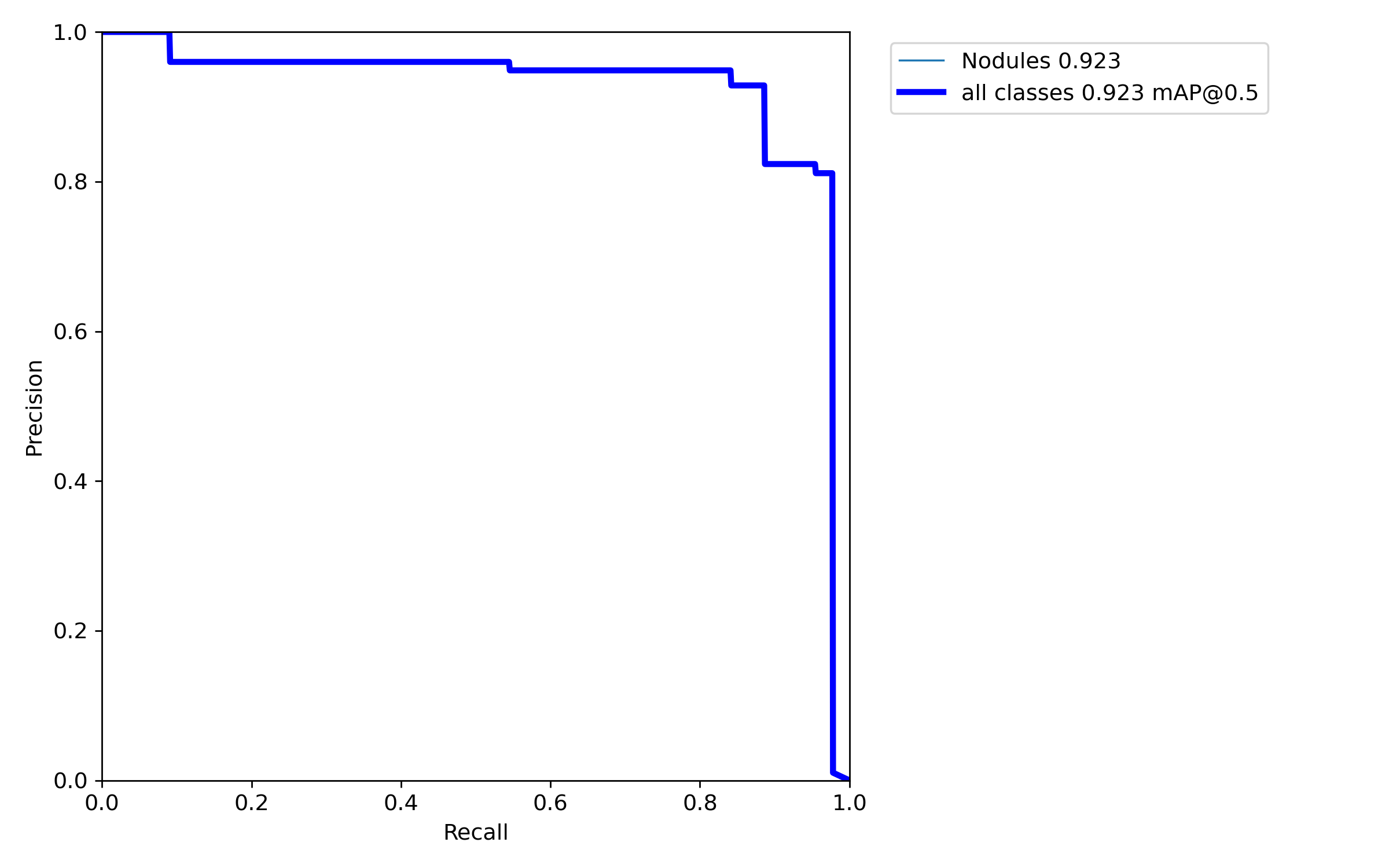}
        \caption{Precision — Recall Curve of the validation data }
        \label{fig:Recall_p}
    \end{subfigure}
    \vspace{1.52ex}
   \caption{the important Evaluation Metrics}
  \label{F1_Score and Precision — Recall Curve for Single Class evaluation }
\end{figure}

To check the model's predictions, and generalizations a few evaluation parameters must be tracked during training and validation. There are several criteria to keep in mind while evaluating a box loss, Precision, and recall values. The variable box benefits from objectivity and categorization. Fig.\ref{fig:plot_tranign} shows all of the graphs that were used for this work. And Figure \ref{fig:F_score} shows the F1 indicator training process for a single category that we want to be detected. The F1 score tends to be 0 with increasing confidence . Training and validation box losses are reduced Fig.\ref{fig:plot_tranign}, suggesting that the model is sound good. the mAP is the abbreviation of median accuracy performances. The high number indicates that this parameter is correct 92.27\% as shown blow in Fig.\ref{fig:Evaluation}.

\vspace{1.52ex}
\pagebreak
Precision is needed to determine how accurate the model forecasts are 92.82\% following the figure \ref{fig:Evaluation}. Only excellent results may be achieved by using the recall method. the model performance showed a good benefit of using Hyper-parameter tuning to make better Learning from data samples and generalize good knowledge from distribution can be seen the following Fig.\ref{fig:plot_tranign} .
Due to the importance of both precision and recall, there is a precision-recall curve the shows the tradeoff between the precision and recall values for different thresholds. This curve helps to select the best threshold to maximize both metrics, tin the following Fig.\ref{fig:Recall_p}
\pagebreak
\section{Discussion Ana Conclusion}
This research examined at how an AI model can help readers detect viewable lung cancer in Ct images. Residents identified more viewable lung cancer when AI was being used as a second reader.In this research, the dataset has been collected from LIDC-IDRI. In LIDC-IDRI image collection, thoracic CT scans with marked-up annotated lesions are included. Yolov5 model is used for feature extraction and detection of lung nodules in CT scans.During the training and validation process, a total of 270 CT Scan images are used of which 239 CT Scans are used for training and 41 are used for validations. In this study, the model's performance was assessed using accuracy, precision, and recall. The accuracy metric indicates how well the model recognised both positive and negative instances. The precision metric measures how well the model predicts both negative and positive cases. The model's high accuracy, precision, and recall imply that it has a small error possibility. Our findings imply that the AI technique assists low experienced individuals in terms of recall while benefiting more-experienced audience in terms of precision. Previous research has revealed that inexperienced readers are more likely to overlook lung malignancies, particularly lesions with a limited visibility score.\vspace{1.5ex}
In this research LIDC-IDRI dataset is used which have lung nodules in it. The purpose of this study  was to identify the nodule that were developing in the lungs of the participants. It was difficult to find information on lung nodules in medical literature. Research in the medical field often use deep learning. Deep learning will be utilised to develop an algorithm with the support of previous medical imaging research, according to the findings of a literature review. Using over 270 CT images , we were able to classify and identify nodules using a deep learning algorithm. Using medical images analysis based on deep neural networks, this study found that as much as 92.27\% of cancer could be detected. Nodules on radiographs are easier to see with its help. Using this technology in the future will help treat illnesses including brain tumours and breast cancer.  

\section{Disclosures}
The authors declare that they have no conflict of interest

\section{Acknowledgments}
We would like to thank our respectful research assistant Moath Alawaqla, for his distinguished role of data collection. 
\section{Funding}
This work supported by Jordan University of Science and Technology, Irbid-Jordan,


\bibliography{report} 
\vspace{2ex}
\bibliographystyle{spiejour}   

\end{document}